\documentclass[journal]{IEEEtran}
\usepackage{amssymb}
\usepackage{amsmath}
\DeclareMathOperator{\tr}{tr}
\usepackage{graphicx}
\usepackage{cite}
\usepackage{txfonts}
\usepackage{mathrsfs}
\usepackage{fontenc}
\usepackage[binary-units]{siunitx}
\usepackage[caption=false,font=footnotesize]{subfig}
\usepackage{tabularx}
\usepackage{multirow}
\usepackage{diagbox}
\usepackage{algorithm}
\usepackage{algorithmic}
\usepackage{bbm}
\usepackage{xcolor}

\begin{document}
\title{Impact of Channel Aging on Zero-Forcing Precoding in Cell-Free Massive MIMO Systems}

\author{Wei~Jiang,~\IEEEmembership{Senior~Member,~IEEE,}
        and Hans~Dieter~Schotten,~\IEEEmembership{Member,~IEEE}
\thanks{\textit{Corresponding author: Wei Jiang (e-mail: wei.jiang@dfki.de)}}
\thanks{W. Jiang and H. D. Schotten are with the Intelligent Networking Research Group, German Research Centre for Artificial Intelligence (DFKI), Kaiserslautern, Germany, and are also with the Department of Electrical and Computer Engineering, Technische University (TU) Kaiserslautern, Germany.}
}

\maketitle

\begin{abstract}
In the context of cell-free massive multi-input multi-output (mMIMO), zero-forcing precoding (ZFP) requires the exchange of instantaneous channel state information and precoded data symbols via a fronthaul network. It causes considerable propagation and processing delays, which degrade performance. This letter analyzes the impact of channel aging on the performance of ZFP in cell-free mMIMO. The aging effects of not only user mobility but also phase noise are considered. Numerical results in terms of per-user spectral efficiency are illustrated.
\end{abstract}

\IEEEpeerreviewmaketitle

\section{Introduction}

\IEEEPARstart{A}{s} the latest advance of distributed massive multi-input multi-output (mMIMO) also known as network MIMO, a recent concept referred to as cell-free mMIMO has received much attention from academia and industry \cite{Ref_ngo2017cellfree}. Applying a large number of transmit antennas makes linear precoding perform nearly as good  as nonlinear precoding such as the optimal dirty-paper coding. Linear precoding is mainly  implemented by conjugate beamforming (CB) or zero-forcing precoding (ZFP). The former is simple with low requirement on backhaul but suffers from inter-user interference. Consequently, many researchers focus on ZFP as well since it can achieve much higher spectral efficiency (SE), regardless of its higher implementation complexity and larger backhaul burden. The superiority of ZFP over CB in terms of spectral efficiency is demonstrated  in \cite{Ref_nayebi2017precoding},  and its energy efficiency is analyzed in \cite{Ref_nguyen2017energy}.  The uplink data throughput of ZFP under imperfect channel state information (CSI) is provided in \cite{Ref_liu2020spectral} and its performance using quantized CSI is given in \cite{Ref_maryopi2019uplink}.

In cell-free mMIMO, ZFP requires the exchange of instantaneous CSI and precoded data between a central processing unit (CPU) and access points (APs) via a fronthaul network (also called backhaul in some literature, so both terms are  interchangeable  hereinafter). Masoumi and Emadi studied the system performance with limited backhaul capacity \cite{Ref_masoumi2020performance}, while Buzzi et al. proposed a user-centric approach to lower the backhaul overhead \cite{Ref_buzzi2020usercentric}. However, these works only focus on the backhaul capacity but do not notice the effect of backhaul delay.  In practice, the performance of ZFP is vulnerable to such a delay since the knowledge of CSI is outdated quickly under the channel fading and imperfect hardware, referred to as channel aging. The impact of channel aging  has been extensively studied in different areas such as collocated mMIMO \cite{Ref_Papazafeiropoulos2017impact, Ref_chopra2018performance}, transmit antenna selection \cite{Ref_WJ_jiang2019neural}, coordinated multi-point transmission \cite{Ref_jaramilloramirez2015coordinated}, and  cooperative diversity \cite{Ref_jiang2016robust, Ref_jiang2021simple}. In cell-free mMIMO, \cite{Ref_zheng2020cellfree} is the unique work  until now but it only focuses on the effect of user mobility on the uplink rate. To the best knowledge of the authors, the downlink performance of cell-free mMIMO in the presence of channel aging is still missing. Meanwhile, the majority of the previous works only consider the Doppler shift raised by the relative mobility of users or surrounding reflectors, whereas the phase noise raised by imperfect oscillator is usually neglected \cite{Ref_Papazafeiropoulos2017impact}.
The main contributions of this letter can be highlighted as follows:
\begin{itemize}
    \item It is the first work to analyze the impact of channel aging on the performance of ZFP in distributed/cell-free mMIMO, and time alignment including all processing and propagation delays over the fronthaul network is modeled.
    \item A closed-form expression of SE, considering the effects of not only user mobility but also phase noise, is derived.
    \item In order to give some insights on the channel aging, representative numerical examples in terms of achievable per-user SE are illustrated.
\end{itemize}

\section{System Model}

We focus on the downlink of a cell-free mMIMO system where $M$ APs are randomly distributed over a geographical area.  The APs are connected to a CPU via a fronthaul network and synchronized to serve $K$ users, where $K\ll M$,  upon the same time-frequency resource. Each AP and user equipment (UE) is equipped with a single antenna.
The downlink (DL) transmission from the APs to the UEs and the uplink (UL) from  the UEs to the APs are separated by time-division duplex with the assumption of perfect channel reciprocity as \cite{Ref_ngo2017cellfree, Ref_nayebi2017precoding}.

\subsection{Channel Model}
We adopt $\sqrt{\beta_{mk}} h_{mk}$ to model the fading channel between AP $m=1,\ldots,M$ and UE $k=1,\ldots,K$, where $\beta_{mk}$ and $h_{mk}$ represent large-scale and small-scale fading, respectively.
Small-scale fading is assumed to be frequency flat and is modelled by a circularly-symmetric complex Gaussian random variable with zero mean and unit variance, i.e., $h_{mk} \sim \mathcal{CN}(0, 1)$. Large-scale fading is frequency independent and keeps constant within a transmission block. It is given by $\beta_{mk}=10^\frac{PL_{mk}+X_{mk}}{10}$ with shadowing fading $X_{mk}\sim \mathcal{N}(0,\sigma_{sd}^2)$ and path loss $PL_{mk}$. As  \cite{Ref_ngo2017cellfree}, the COST-Hata model is applied, i.e.,
\begin{equation} \label{eqn:CostHataModel}
    PL_{mk}=
\begin{cases}
-L-35\log_{10}(d_{mk}), &  d_{mk}>d_1 \\
-L-10\log_{10}(d_1^{1.5}d_{mk}^2), &  d_0<d_{mk}\leq d_1 \\
-L-10\log_{10}(d_1^{1.5}d_0^2), &  d_{mk}\leq d_0
\end{cases},
\end{equation}
where $d_{mk}$ represents the distance between AP $m$ and UE $k$,  $d_0$ and $d_1$ are the  break points of the three-slope model, and
\begin{IEEEeqnarray}{ll}
 L=46.3&+33.9\log_{10}\left(f_c\right)-13.82\log_{10}\left(h_{AP}\right)\\ \nonumber
 &-\left[1.1\log_{10}(f_c)-0.7\right]h_{UE}+1.56\log_{10}\left(f_c\right)-0.8
\end{IEEEeqnarray} with carrier frequency $f_c$, the antenna height of AP $h_{AP}$, and the antenna height of UE $h_{UE}$.

\subsection{Channel Aging}
The performance of ZFP is affected heavily by the quality of instantaneous CSI \cite{Ref_liu2020spectral, Ref_maryopi2019uplink}. In cell-free mMIMO, there is a time gap between the instant when pilot sequences sound the UL channels and the instant of the DL data transmission due to the processing and propagation delays over the fronthaul network.
Due to \textit{user mobility} and \textit{phase noise}, the acquired CSI may be outdated. This is called channel aging \cite{Ref_Papazafeiropoulos2017impact, Ref_chopra2018performance}, which may degrade the system performance substantially.
\subsubsection{User Mobility}
The relative movement between an AP and a UE as well as their surrounding reflectors leads to a time-varying channel. Given the moving speed $v_k$ of a typical UE $k$, its maximal Doppler shift is obtained by $f_d^k=v_k/\lambda$, where $\lambda$ represents the wavelength of carrier frequency. The higher the mobility, the faster the channel varies. To quantify the aging of CSI raised by the Doppler effect, a metric known as correlation coefficient is applied, as defined by \cite{Ref_jiang2021simple}
\begin{equation}
\label{Eqn_CorCoeff}
\rho_k=\frac{\mathbb{E}\left[h_{mk,d}h_{mk,p}^*\right]}{\sqrt{\mathbb{E}[|h_{mk,p}|^2] \mathbb{E}[|h_{mk,d}|^2]}},
\end{equation}
where $\mathbb{E}\left[ \cdot \right]$ stands for mathematical expectation, $h_{mk,p}$ and $h_{mk,d}$ denote the small-scale channel fading between AP $m$ and UE $k$ at the instants of the UL training (notated by $p$) and DL data transmission (notated by $d$), respectively.  Under the classical Doppler spectrum of the Jakes model, it takes the value $\rho_k=J_0(2\pi f_d^k \triangle\tau)$, where $\triangle\tau$ stands for the overall delay, and $J_0(\cdot)$ denotes the $zeroth$ order  Bessel  function  of the first kind. According to \cite{Ref_jiang2016robust}, we have
\begin{equation} \label{eqn:outdatedCSI}
h_{mk,d}=\left ( \rho_k h_{mk,p} + \varepsilon_{mk} \sqrt{1-\rho_k^2}  \right )
\end{equation}
with an innovation component $\varepsilon_{mk}$ that is a random variable with standard normal distribution $\varepsilon_{mk} \sim \mathcal{CN}(0,1)$.
\subsubsection{Phase Noise}
It is attractive for cost-efficient implementation of mMIMO systems with low-cost transceivers, but the problem of hardware impairments raises. Meanwhile, each distributed AP in cell-free mMIMO has to operate a local oscillator, in contrast to a common oscillator in a collocated mMIMO setup.  Due to imperfect oscillators at the transmitter, the generated signals suffer from phase noise during the up-conversion from baseband to passband signals, and \textit{vice versa} at the receiver \cite{Ref_Papazafeiropoulos2017impact}. Such phase noise is not only random but also time-varying \cite{Ref_ozdogan2019performance}, leading to the outdated CSI that is equivalent to that of user mobility. However, this effect is usually neglected in previous works such as \cite{Ref_chopra2018performance,  Ref_jaramilloramirez2015coordinated, Ref_jiang2016robust, Ref_jiang2021simple}.

Utilizing a well-established  Wiener process \cite{Ref_Papazafeiropoulos2017impact, Ref_krishnan2016linear}, the phase noise of the $m^{th}$ AP and the $k^{th}$ UE at \textit{discrete-time} instant $t$ can be modeled as
\begin{align} \label{phaseNoiseBS} \nonumber
 \phi_{m,t}&=\phi_{m,t-1}+\triangle \phi_{t}, &\triangle \phi_{t}\sim \mathcal{CN}(0,\sigma_{\phi_{}}^{2}) \\
 \varphi_{k,t}&=\varphi_{k,t-1}+\triangle \varphi_{t}, &\triangle \varphi_{t} \sim \mathcal{CN}(0,\sigma_{\varphi_{}}^{2}),
\end{align}
where the increment variances are given by $\sigma_{i}^{2}=4\pi^{2}f_{\mathrm{c}} c_{i}T_{\mathrm{s}}$, $i=\phi, \varphi$ with  symbol period $T_{\mathrm{s}}$ and oscillator-dependent constant  $c_{{i}}$.
Then, we can write \begin{equation} \label{Eqn_ChannelModel}
g_{mk,t}=\sqrt{\beta_{mk}} h_{mk,t} e^{j(\phi_{m,t}+\varphi_{k,t})}
\end{equation} to denote the overall channel gain between AP $m$ and UE $k$ at instant $t$ combining the effects of path loss, shadowing, small-scale fading, and phase noise. In particular, the acquired CSI $g_{mk,p}=\sqrt{\beta_{mk}} h_{mk,p} e^{j(\phi_{m,p}+\varphi_{k,p})}$ is an outdated version of its actual value $g_{mk,d}=\sqrt{\beta_{mk}} h_{mk,d} e^{j(\phi_{m,d}+\varphi_{k,d})}$. Under good condition where the channels exhibit slow fading under low mobility and the quality of oscillators is high, the effect of channel aging is not explicit and the performance loss is possible to be small. Otherwise, the impact should be analyzed seriously.

\section{The Communications Process}
\begin{figure}[!t]
    \centering
    \includegraphics[width=0.49\textwidth]{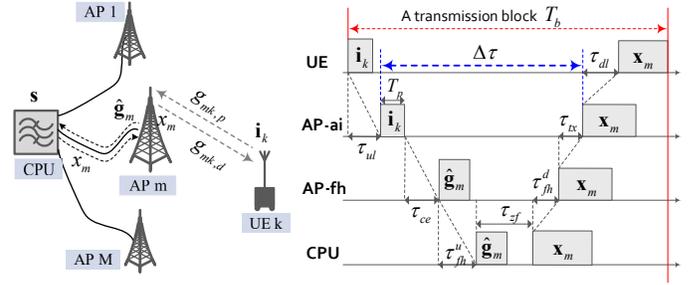}
    \caption{Schematic diagram and time alignment of ZFP in a cell-free mMIMO system where $M$ APs serve $K$ UEs. When UE $k$ sends the pilot sequence $\mathbf{i}_k$, the instantaneous CSI is $g_{mk,p}$. Then, AP $m$ estimates and transfers $\hat{\mathbf{g}}_m$ to the CPU such that $\mathbf{s}$ can be precoded. Due to the delay $\triangle \tau$, the CSI during the DL data transmission changes, namely $g_{mk,d}\neq g_{mk,p}$. The data block for AP $m$ is denoted by $\mathbf{x}_m=[x_{m,1},\ldots,x_{m,N}]^T$, where $N$ is the number of precoded symbols per block. Note that \textit{AP-ai} stands for the air interface of an AP while \textit{AP-fh} points to the part that interacts with the fronthaul network.  }
    \label{fig:zeroforcingprecoding}
\end{figure}
This section introduces the communication process of a cell-free mMIMO system using zero-forcing precoding.  A transmission block is divided into two phases: the UL training and DL data transmission, while the UL data transmission is neglected to focus on the DL performance analysis. Assume that the APs and the UEs are well-synchronized, and the knowledge of $\beta_{mk}$ is perfectly available. We also assume that the fronthaul network provides error-free and infinite capacity following the setup of previous works such as \cite{Ref_ngo2017cellfree, Ref_nayebi2017precoding, Ref_nguyen2017energy, Ref_liu2020spectral} so as to particularly focus on the backhaul delay. Hence, the discussion of practical backhaul constraints is out of scope for this letter and can refer to \cite{Ref_maryopi2019uplink, Ref_masoumi2020performance, Ref_buzzi2020usercentric}.
As illustrated in \figurename \ref{fig:zeroforcingprecoding}, the system operates as follows:
\begin{itemize}
\item UEs transmit their pilot sequences $\mathbf{i}_k$, $k=1,\ldots,K$ with a duration of $T_p$. The propagation delay is  $\tau_{ul}$.
\item The $m^{th}$ AP, where $m=1,\ldots, M$, can get the estimates $\hat{g}_{mk,p}$, $k=1,\ldots,K$ with a processing time of $\tau_{ce}$.
\item AP $m$ sends its local CSI $\hat{\mathbf{g}}_{m}=\left[\hat{g}_{m1,p},\ldots,\hat{g}_{mK,p}\right]^T\in \mathbb{C}^{K\times 1}$  to the CPU, leading to a propagation delay of $\tau_{fh}^u$.
\item Using $\hat{\mathbf{G}}=[\hat{\mathbf{g}}_{1},\ldots,\hat{\mathbf{g}}_{M}] \in \mathbb{C}^{K\times M}$, the CPU precodes the transmit symbols $\mathbf{s}=[s_1,\ldots,s_K]^T$, where $\mathbb{E}[\vert s_k \vert^2]=1$. The time cost of the precoding is $\tau_{zf}$.
\item The CPU distributes the precoded symbol $x_m$ to   AP $m$, using the time of $\tau_{fh}^d$.
\item The transmitter of AP $m$ needs a preparation time of $\tau_{tx}$ to start the transmission after the reception of $x_m$ and  the propagation of the signal takes $\tau_{dl}$.
\end{itemize}
In particular, let $\triangle \tau$ denote the gap between the time when the pilot sequences probe the channels and the instant that all APs synchronously transmit the precoded symbols. As shown in \figurename \ref{fig:zeroforcingprecoding}, we get $\triangle \tau=T_p+\tau_{ce}+\tau_{fh}^u+\tau_{zf}+\tau_{fh}^d+\tau_{tx}$, which is normalized by the sampling period to $n_{\triangle \tau}=\left \lceil \frac{ \triangle \tau}{T_s} \right \rceil$.

\subsection{Uplink Training}

\begin{figure*}[!t]
\setcounter{equation}{15}
\begin{align} \label{eqn:longZFRx} \nonumber
y_k &= \sqrt{p_d} \mathbf{g}_{k,d} \hat{\mathbf{G}}^H\left(\hat{\mathbf{G}}\hat{\mathbf{G}}^H\right)^{-1} \mathbf{P}\mathbf{s}+z_k\\
                &= \sqrt{p_d} e^{j\left(\varphi_{k,d}-\varphi_{k,p}\right)} \left( \rho_k\hat{\mathbf{g}}_k +\rho_k \tilde{\mathbf{g}}_k +\sqrt{1-\rho_k^2} \mathbf{e}_k   \right) \triangle \boldsymbol{ \Phi} \hat{\mathbf{G}}^H\left(\hat{\mathbf{G}}\hat{\mathbf{G}}^H\right)^{-1} \mathbf{P}\mathbf{s}+z_k\\ \nonumber
                 &= \underbrace{ \sqrt{p_d \eta_k} e^{j\left(\varphi_{k,d}-\varphi_{k,p}\right)} e^{- \frac{n_{ \triangle \tau}\sigma_{\phi}^2}{2}} \rho_ks_k}_{\mathrm{Desired\:Signal}:\: \mathcal{D}}+\underbrace{\sqrt{p_d} e^{j(\varphi_{k,d}-\varphi_{k,p})} \rho_k  \tilde{\mathbf{g}}_k \triangle \boldsymbol{ \Phi} \hat{\mathbf{G}}^H\left(\hat{\mathbf{G}}\hat{\mathbf{G}}^H\right)^{-1} \mathbf{P}\mathbf{s} }_{\mathrm{Effective\:Noise:} \:\mathcal{I}_1} + \underbrace{ \sqrt{p_d(1-\rho_k^2)} e^{j(\varphi_{k,d}-\varphi_{k,p})} \mathbf{e}_k  \triangle \boldsymbol{ \Phi}   \hat{\mathbf{G}}^H\left(\hat{\mathbf{G}}\hat{\mathbf{G}}^H\right)^{-1} \mathbf{P}\mathbf{s}}_{\mathrm{Effective\:Noise:}\:\mathcal{I}_2} +\underbrace{z_k}_{\mathcal{I}_3}
\end{align} \rule{\textwidth}{0.1mm}
\begin{align} \label{eqn:I1}  \nonumber
   \mathbb{E}\left[\vert \mathcal{I}_1 \vert^2\right]&=\mathbb{E}\left[\left \vert \sqrt{p_d} e^{j\left(\varphi_{k,d}-\varphi_{k,p}\right)}  \rho_k \tilde{\mathbf{g}}_k    \triangle \boldsymbol{ \Phi} \hat{\mathbf{G}}^H\left(\hat{\mathbf{G}}\hat{\mathbf{G}}^H\right)^{-1} \mathbf{P}\mathbf{s} \right \vert^2 \right] \\
   &= p_d\rho_k^2  \mathbb{E}\left[\left \vert  \tilde{\mathbf{g}}_k\triangle \boldsymbol{ \Phi}\hat{\mathbf{G}}^H\left(\hat{\mathbf{G}}\hat{\mathbf{G}}^H\right)^{-1} \mathbf{P}\mathbf{s} \right \vert^2 \right]=  p_d\rho_k^2 e^{- n_{\triangle \tau}\sigma_{\phi}^2}\tr \left\{ \mathbf{P}^2\mathbb{E}\left[  \left(\hat{\mathbf{G}}\hat{\mathbf{G}}^H\right)^{-1}\hat{\mathbf{G}}   \mathbb{E}\left( \tilde{\mathbf{g}}_k^H \tilde{\mathbf{g}}_k \right)  \hat{\mathbf{G}}^H\left(\hat{\mathbf{G}}\hat{\mathbf{G}}^H\right)^{-1} \right]\right\}\\
   &= \nonumber p_d\rho_k^2 e^{- n_{\triangle \tau}\sigma_{\phi}^2} \tr \left\{ \mathbf{P}^2\mathbb{E}\left[  \left(\hat{\mathbf{G}}\hat{\mathbf{G}}^H\right)^{-1}\hat{\mathbf{G}}   \mathbf{D}_k  \hat{\mathbf{G}}^H\left(\hat{\mathbf{G}}\hat{\mathbf{G}}^H\right)^{-1} \right]\right\}= p_d\rho_k^2 e^{- n_{\triangle \tau}\sigma_{\phi}^2} \sum_{i=1}^K \eta_{i}\xi_{ki}
\end{align} \rule{\textwidth}{0.1mm}
\begin{align} \label{eqn:I2}
   \mathbb{E}\left[\vert \mathcal{I}_2 \vert^2\right]&=\mathbb{E}\left[\left \vert \sqrt{p_d(1-\rho_k^2)} e^{j\left(\varphi_{k,d}-\varphi_{k,p}\right)} \mathbf{e}_k \triangle \boldsymbol{ \Phi}  \hat{\mathbf{G}}^H\left(\hat{\mathbf{G}}\hat{\mathbf{G}}^H\right)^{-1} \mathbf{P}\mathbf{s} \right \vert^2 \right] = p_d\left(1-\rho_k^2\right)  \mathbb{E}\left[\left \vert  \mathbf{e}_k \triangle \boldsymbol{ \Phi} \hat{\mathbf{G}}^H\left(\hat{\mathbf{G}}\hat{\mathbf{G}}^H\right)^{-1} \mathbf{P}\mathbf{s} \right \vert^2 \right]\\
    &= \nonumber p_d \left(1-\rho_k^2\right) e^{- n_{\triangle \tau}\sigma_{\phi}^2} \tr \left\{ \mathbf{P}^2\mathbb{E}\left[  \left(\hat{\mathbf{G}}\hat{\mathbf{G}}^H\right)^{-1}\hat{\mathbf{G}}   \mathbf{E}_k  \hat{\mathbf{G}}^H\left(\hat{\mathbf{G}}\hat{\mathbf{G}}^H\right)^{-1} \right]\right\}=  p_d \left(1-\rho_k^2\right) e^{- n_{\triangle \tau}\sigma_{\phi}^2} \sum_{i=1}^K \eta_{i}\chi_{ki}
\end{align} \rule{\textwidth}{0.1mm}
\end{figure*}
During the training phase, the channels and phase noise are assumed to remain constant to yield a simple model. This is a valid assumption since the duration of the training phase scales with the number of users, which is small.
Let the length of pilot sequences be $\tau_p$ symbols, corresponding to $T_p=\tau_p T_s$. There are a total of $\tau_p$ orthogonal pilot sequences, which are represented by  $\boldsymbol \Omega = [\boldsymbol \omega_{1}, \ldots, \boldsymbol \omega_{\tau_p}] \in \mathbb{C}^{\tau_p \times \tau_p}$ with $\boldsymbol \Omega$ normalized, i.e., $\boldsymbol \Omega \boldsymbol \Omega^H = \mathbf{I}_{\tau_p}$.
Denoting the pilot sequence of UE $k$ by $\mathbf{i}_k$, we have $\exists ! x{\in}\{1,\ldots,\tau_p\}\: \left( \mathbf{i}_k = \boldsymbol \omega_{x}\right)$, where $\exists !$ stands for unique existential quantification.  UEs simultaneously transmit their  pilot sequences, resulting in an observation at AP $m$ as
\setcounter{equation}{6}
\begin{equation}
    \mathbf{y}_{m,p} = \sqrt{p_u} \sum_{k=1}^K g_{mk,p} \mathbf{i}_k+\mathbf{z}_{m,p},
\end{equation}
where $p_u$ is the transmit power limit of each UE and $\mathbf{z}$ represents additive noise with variance $\sigma_z^2$.
Correlating $\mathbf{y}_{m,p}$ with the known pilot sequence, yields
\begin{equation}
y_{mk,p} =\mathbf{i}_k^H \mathbf{y}_{m,p}= \sqrt{p_u} g_{mk,p} +  \sqrt{p_u} \sum_{k'\neq k}^K g_{mk,p} \mathbf{i}_k^H \mathbf{i}_{k'}+ \mathbf{i}_k^H \mathbf{z}_{m,p}.
\end{equation}
Owing to the limitation of the frame length, some UEs need to share the same sequence if $\tau_p<K$, leading to pilot contamination. We use $\mathcal{P}_k$ to denote the set of indices for the users, including user $k$, that utilize the same sequence as $\mathbf{i}_k$. Let $\hat{g}_{mk,p}$ be an estimate of $g_{mk,p}$ and $\tilde{g}_{mk,p}$ be the estimation error raised by additive noise and pilot contamination, we have
\begin{equation} \label{eqn:ChannelEstiError}
    \hat{g}_{mk,p} = g_{mk,p} - \tilde{g}_{mk,p}.
\end{equation}
Conducting channel estimation with linear minimum mean-square error (MMSE) \cite{Ref_ngo2017cellfree}, the estimate is obtained by
\begin{equation} \nonumber
    \hat{g}_{mk,p} = \frac{\mathbb{E}\left[  y_{mk,p}^* g_{mk,p}\right] y_{mk,p}}{\mathbb{E}\left[ \left \vert y_{mk,p}\right \vert^2\right]} = \left(\frac{\sqrt{p_u}\beta_{mk}}{p_u\sum_{k'\in \mathcal{P}_k } \beta_{mk'} + \sigma_z^2}\right)y_{mk,p}.
\end{equation}
Other methods such as zero-forcing estimation can also be applied to acquire CSI, but it is independent with zero-forcing precoding discussed in this letter.
The variance of $\hat{g}_{mk,p}$ is
\begin{equation}
\mathbb{E}\left[\left \vert \hat{g}_{mk,p} \right \vert ^2 \right]=\frac{p_u\beta_{mk}^2}{p_u\sum_{k'\in \mathcal{P}_k } \beta_{mk'} + \sigma_z^2}.
\end{equation}
Then, we know that $\hat{g}_{mk,p}\in \mathcal{CN}(0,\alpha_{mk})$ with $\alpha_{mk}=\frac{p_u\beta_{mk}^2}{p_u \sum_{k'\in \mathcal{P}_k } \beta_{mk'} + \sigma_z^2}$ and $\tilde{g}_{mk,p}\in \mathcal{CN}(0,\beta_{mk}-\alpha_{mk})$, in comparison with the actual CSI $g_{mk,p}\in \mathcal{CN}(0,\beta_{mk})$.

\subsection{Downlink Data Transmission}
Without loss of generality, we assume that CSI keeps constant during the data transmission.
The real downlink channel matrix is denoted by $\mathbf{G} \in \mathbb{C}^{K\times M}$, where the entry on the $k^{th}$ row and the $m^{th}$ column is $g_{km,d}=g_{mk,d}$ due to channel reciprocity. But the system only knows  $\hat{\mathbf{G}}$ consisting of $\hat{g}_{mk,p}$ that is not only an estimate but also an aged version of $\mathbf{G}$.  With the precoded symbol vector $\mathbf{x}=\left[x_1,\ldots,x_M \right]^T$, the received symbol vector $\mathbf{y}=\left[y_1,\ldots,y_K \right]^T$, and the transmit power limit of AP $p_d$, the DL data transmission is expressed by
\begin{equation}
    \label{eqn:RxSignal}\mathbf{y} =\sqrt{\rho_d} \mathbf{G}\mathbf{x}+\mathbf{z}.
\end{equation}
Using ZFP, we get the precoded symbol vector as $\mathbf{x}=\hat{\mathbf{G}}^H\left(\hat{\mathbf{G}}\hat{\mathbf{G}}^H\right)^{-1}\mathbf{P}\mathbf{s}$, where $\mathbf{P}\in \mathbb{C}^{K\times K}$ is a diagonal matrix consisting of power-control coefficients, i.e., $\mathbf{P}=\mathrm{diag}\{\eta_1,\ldots,\eta_K\}$. Then, (\ref{eqn:RxSignal}) can be rewritten as
\begin{equation} \label{eqn:RxSignal_matrix}
\mathbf{y} = \sqrt{\rho_d}\mathbf{G} \hat{\mathbf{G}}^H\left(\hat{\mathbf{G}}\hat{\mathbf{G}}^H\right)^{-1} \mathbf{P}\mathbf{s}+\mathbf{z}.
\end{equation}

\section{Performance Analysis}
According to (\ref{Eqn_ChannelModel}), the overall CSI during the DL data transmission is obtained:
\begin{equation} \label{eqn:overallCSI}
    g_{mk,d} =\sqrt{\beta_{mk}} h_{mk,d} e^{j(\phi_{m,d}+\varphi_{k,d})}.
\end{equation}
The innovation component in (\ref{eqn:outdatedCSI}) corresponds to a compound element in the overall CSI, which we write as $e_{mk}=\sqrt{\beta_{mk}}\varepsilon_{mk} e^{j(\phi_{m,p}+\varphi_{k,p})}$ for ease of derivation.
Substituting (\ref{eqn:outdatedCSI}) into (\ref{eqn:overallCSI}) and applying (\ref{eqn:ChannelEstiError}), we have
\begin{IEEEeqnarray} {lll} \nonumber \label{eqn:OVERallCSIdecompose}
   g_{mk,d} &=&\sqrt{\beta_{mk}} \left(\rho_k h_{mk,p} + \varepsilon_{mk}\sqrt{1-\rho_k^2} \right)e^{j\left(\phi_{m,d}+\varphi_{k,d}+\phi_{m,p}+\varphi_{k,p}-\phi_{m,p}-\varphi_{k,p}\right)}\\    &=& \left(\rho_k g_{mk,p} + e_{mk}\sqrt{1-\rho_k^2}\right) e^{j\left(\phi_{m,d}+\varphi_{k,d}-\phi_{m,p}-\varphi_{k,p}\right)}\\ \nonumber
            &=& \left(\rho_k \hat{g}_{mk,p} + \rho_k \tilde{g}_{mk,p} +e_{mk}\sqrt{1-\rho_k^2}\right) e^{j\left(\phi_{m,d}-\phi_{m,p}\right)} e^{j\left(\varphi_{k,d}-\varphi_{k,p}\right)} .
\end{IEEEeqnarray}
We denote the $k^{th}$ row of $\hat{\mathbf{G}}$ as $\hat{\mathbf{g}}_{k}=\left[\hat{g}_{1k,p},\ldots,\hat{g}_{Mk,p}\right]$, while defining  $\tilde{\mathbf{g}}_{k}=\left[\tilde{g}_{1k,p},\ldots,\tilde{g}_{Mk,p}\right]$, $\mathbf{e}_{k}=\left[e_{1k},\ldots,e_{Mk}\right]$, and a diagonal matrix $\triangle \boldsymbol{\Phi}= \mathrm{diag}\{e^{j(\phi_{1,d}-\phi_{1,p})},\ldots,e^{j(\phi_{M,d}-\phi_{M,p})} \}\in \mathbb{C}^{M\times M}$. Building a channel vector $\mathbf{g}_{k,d}=\left[g_{1k,d},\ldots,g_{Mk,d}\right]\in \mathbb{C}^{1\times M}$ and substituting (\ref{eqn:OVERallCSIdecompose}) into it, yields
\begin{equation} \label{eqn:CSIvectorRx}
   \mathbf{g}_{k,d} = e^{j\left(\varphi_{k,d}-\varphi_{k,p}\right)} \left( \rho_k\hat{\mathbf{g}}_k +\rho_k \tilde{\mathbf{g}}_k +\sqrt{1-\rho_k^2} \mathbf{e}_k   \right) \triangle \boldsymbol{ \Phi}
\end{equation} with several steps manipulation.
From (\ref{eqn:RxSignal_matrix}), we can get the received signal of UE $k$, i.e., $y_k = \sqrt{p_d} \mathbf{g}_{k,d} \hat{\mathbf{G}}^H\left(\hat{\mathbf{G}}\hat{\mathbf{G}}^H\right)^{-1} \mathbf{P}\mathbf{s}+z_k$, like \cite{Ref_nayebi2017precoding}. Using (\ref{eqn:CSIvectorRx}),  the received signal can be decomposed, as given by (\ref{eqn:longZFRx}). During the derivation, we apply  $T_{PN}=\lim_{M\rightarrow \infty} \frac{1}{M} \mathrm{tr}\left \{ \triangle \boldsymbol{\Phi} \right \} = e^{- n_{\triangle \tau}\sigma_{\phi}^2/2}$ according to \cite{Ref_krishnan2016linear}, where the phase noise hardens to a deterministic value when $M\rightarrow \infty$.

The transmit symbols, estimation errors, innovation components, and additive noise are independent, such that the terms $\mathcal{D}$, $\mathcal{I}_1$, $\mathcal{I}_2$, and $\mathcal{I}_3$ in (\ref{eqn:longZFRx}) are mutually uncorrelated. By using the fact that uncorrected Gaussian noise represents the worst case \cite{Ref_ngo2017cellfree}, the achievable SE is lower bounded by $\log_2\left( 1+\gamma_{k}\right)$ with the effective signal-to-interference-plus-noise ratio (SINR)
\setcounter{equation}{18}
\begin{equation} \label{eqn:SNRcompute}
    \gamma_{k}=  \frac{\mathbb{E}[\vert \mathcal{D} \vert^2]}{\mathbb{E}[\vert \mathcal{I}_1 \vert^2] + \mathbb{E}[\vert \mathcal{I}_2 \vert^2] +\mathbb{E}[\vert \mathcal{I}_3 \vert^2] }.
\end{equation}
It is easy to know $\mathbb{E}[\vert \mathcal{D} \vert^2]=p_d\eta_k\rho_k^2 e^{- n_{\triangle \tau}\sigma_{\phi}^2}$ and $\mathbb{E}[\vert \mathcal{I}_3 \vert^2]=\sigma_z^2$.
Similar to \textit{Theorem 1} of \cite{Ref_nayebi2017precoding}, we can figure out $\mathbb{E}\left[\vert \mathcal{I}_1 \vert^2\right]$, as given by (\ref{eqn:I1}) at the top of this page,
where $\mathbf{D}_k=\mathbb{E}\left( \tilde{\mathbf{g}}_k^H \tilde{\mathbf{g}}_k \right)=\mathrm{diag}\left(\beta_{1k}-\alpha_{1k},\beta_{2k}-\alpha_{2k},\ldots,\beta_{Mk}-\alpha_{Mk}\right)\in \mathcal{C}^{M\times M}$ is a diagonal matrix, and $\xi_{ki}$ denotes the $i^{th}$ diagonal element of $ \mathbb{E}\left[  \left(\hat{\mathbf{G}}\hat{\mathbf{G}}^H\right)^{-1}\hat{\mathbf{G}}   \mathbf{D}_k  \hat{\mathbf{G}}^H\left(\hat{\mathbf{G}}\hat{\mathbf{G}}^H\right)^{-1} \right] $. Likewise, $\mathbb{E}\left[\vert \mathcal{I}_2 \vert^2\right]$ is given by (\ref{eqn:I2}),
where $\mathbf{E}_k=\mathbb{E}\left( \mathbf{e}_k^H \mathbf{e}_k \right)=\mathrm{diag}\left(\beta_{1k},\beta_{2k},\cdots,\beta_{Mk}\right)\in \mathcal{C}^{M\times M}$, and $\chi_{ki}$ represents the $i^{th}$ diagonal element of $ \mathbb{E}\left[  \left(\hat{\mathbf{G}}\hat{\mathbf{G}}^H\right)^{-1}\hat{\mathbf{G}}   \mathbf{E}_k  \hat{\mathbf{G}}^H\left(\hat{\mathbf{G}}\hat{\mathbf{G}}^H\right)^{-1} \right] $.
Substituting (\ref{eqn:I1}) and (\ref{eqn:I2}) into (\ref{eqn:SNRcompute}), we get
\begin{equation}
   \gamma_{k}
    =  \frac{ \rho_k^2 \eta_k }
    {\rho_k^2 \sum_{i=1}^{K}\eta_{i} \xi_{ki}+ \left(1-\rho_k^2\right) \sum_{i=1}^{K}\eta_{i} \chi_{ki}+ \frac{\sigma^2_z}{p_d e^{- n_{\triangle \tau}\sigma_{\phi}^2}}}.
\end{equation}
Taking into account the propagation delay over the air interface $n_{ai}=\left \lceil \frac{\tau_{ul}+\tau_{dl}}{T_s} \right \rceil$ and the delay $n_{\triangle \tau}$, the achievable spectral efficiency of the $k^{th}$ user is
\begin{equation}
    R_{k}= \left( 1-\frac{n_{ai}+n_{\triangle \tau}}{T_b} \right)   \log_2\left(1+\gamma_{k}\right).
\end{equation}

\section{Numerical Examples}

\begin{figure}[!t]
    \centering
    \includegraphics[width=0.32\textwidth]{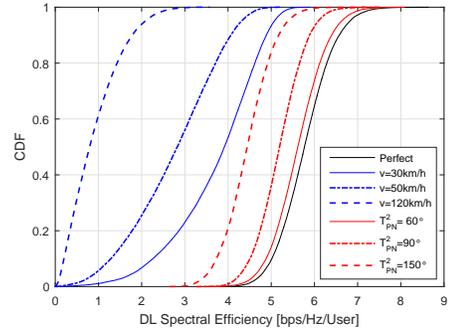}
   \caption{The CDF comparison under different values of user mobility and phase noise. The curve of perfect CSI in stationary state $v=0\mathrm{km/h}$ with perfect oscillators $T_{PN}=0^\circ$ is applied as the benchmark.}
    \label{fig:resultsCDF}
\end{figure}
\begin{figure}[!t]
    \centering
   \includegraphics[width=0.33\textwidth]{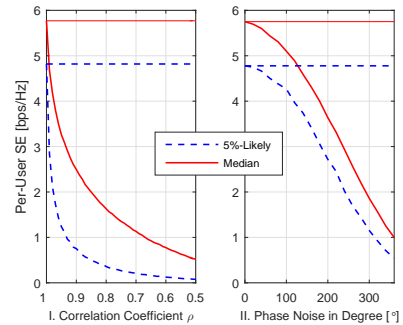}
    \caption{The $5\%$-likely and median per-user SE with respect to correlation coefficient (left) and phase noise (right). The straight lines are $5\%$-likely and median per-user SE of the benchmark, respectively.}
\label{fig:resultsUMandPN}
\end{figure}

This section illustrates some representative numerical results  to observe the impact of channel aging  on the downlink of a cell-free mMIMO system. Note that it is not strictly real-sense simulation but some examples that are got by plugging in some parametric values based on the derived formulae as some previous works \cite{Ref_nayebi2017precoding, Ref_nguyen2017energy}. The achievable per-user SE of ZFP in the presence of user mobility and phase noise serves as the metric.
Consider a square dense urban area of $1\times 1\mathrm{km^2}$ where $M=128$ distributed APs serve $K=16$ single-antenna UEs at the same time-frequency resource. The break points of the three-slope model in (\ref{eqn:CostHataModel}) take values $d_0=10\mathrm{m}$ and $d_1=50\mathrm{m}$. The quantity $L=140.72\mathrm{dB}$ with $f_c=1.9\mathrm{GHz}$, $h_{AP}=15\mathrm{m}$, and $h_{UE}=1.65\mathrm{m}$, while the standard derivation for shadowing fading is $\sigma_{sd}=8\mathrm{dB}$.
The maximum transmit power of AP and UE are $p_d=0.2\mathrm{W}$ and $p_u=0.1\mathrm{W}$, respectively. Due to high complexity of the optimal max-min power control, the APs adopt a sub-optimal scheme with low complexity given in \cite{Ref_nayebi2017precoding}.  That is, $\eta_{1}=\ldots=\eta_{K}=\left( \max_m  \sum_{k=1}^{K} \delta_{km} \right)^{-1}$, where $\boldsymbol \delta_m= \left[\delta_{1m},\ldots,\delta_{Km}\right]^T=\mathrm{diag}\left(\mathbb{E}\left[  \left(\hat{\mathbf{G}}\hat{\mathbf{G}}^H\right)^{-1}    \hat{\mathbf{g}}_m \hat{\mathbf{g}}_m^H   \left(\hat{\mathbf{G}}\hat{\mathbf{G}}^H\right)^{-1} \right]\right)$ and $\hat{\mathbf{g}}_m$ denotes the $m^{th}$ column of $\hat{\mathbf{G}}$.  The variance of white noise  is calculated by $\sigma_z^2=\kappa\cdot B\cdot T_0\cdot N_f$ with the Boltzmann constant $\kappa$, signal bandwidth $B=20\mathrm{MHz}$, temperature $T_0=290 \mathrm{Kelvin}$, and  noise figure  $N_f=9\mathrm{dB}$. Last but not least, the total overhead of the block transmission is set to $(n_{ai}+n_{\triangle \tau})/T_b=10\%$.

\figurename \ref{fig:resultsCDF} provides an comparison with respect to cumulative distribution functions (CDFs) of per-user SE by varying the velocity $v$ or accumulative phase noise $T_{PN}$. The performance curve of ZFP using perfect CSI is applied as a benchmark, where the UEs are stationary ($v=0\mathrm{km/h}$) and the transceivers have perfect local oscillators ($T_{PN}=0^\circ$). To observe the effect of user mobility, we first set $T_{PN}=0^\circ$ and select three typical values: $v=30\mathrm{km/h}$, $50\mathrm{km/h}$, and $120\mathrm{km/h}$. Without loss of generality, the overall delay is simply set to $\triangle \tau=1\mathrm{ms}$ since the aging effect of user mobility is decided by the combination of velocity and delay.  Even at low mobility of $v=30\mathrm{km/h}$, which is equivalent to very high correlation $\rho = 0.97$, the performance deterioration is already remarkable.  To be specific, the $5\%$-likely per-user SE reduces to $1.8\mathrm{bps/Hz}$, in comparison with $4.8\mathrm{bps/Hz}$ of the benchmark, amounting to a loss of $62.5\%$. The $50\%$-likely (median) per-user SE degrades $32\%$, dropping from $5.7\mathrm{bps/Hz}$ to $3.9\mathrm{bps/Hz}$.   With the increase of $v$, the performance loss becomes more substantial. At high mobility of $v=120\mathrm{km/h}$, the $5\%$-likely and median SE further decrease to $0.13\mathrm{bps/Hz}$ and $0.79\mathrm{bps/Hz}$, amounting to a very high loss of $97\%$ and $86\%$, respectively.
In addition, the impact of phase noise is investigated by using the selected phase noise of $T_{PN}^2=30^\circ$, $90^\circ$, and $150^\circ$, where the UEs are set to be stationary $v=0\mathrm{km/h}$. With small phase noise of $30^\circ$, as shown in the figure, the performance loss is marginal. Increased to $150^\circ$, the $5\%$-likely and median SE degrade to $3.5\mathrm{bps/Hz}$ and $4.5\mathrm{bps/Hz}$, equivalent to a loss of $27\%$ and $23\%$, respectively.

Furthermore, \figurename \ref{fig:resultsUMandPN} shows the $5\%$-likely and median per-user SE with respect to correlation coefficient in part I and phase noise in part II. It implies that the system is sensitive to user mobility, where a small decrease of $\rho$ from perfect CSI raises shape  degradation and the performance curve drops to a floor quickly. In contrast, the effect of phase noise is relatively mild and the loss only becomes evident with large phase noise.  Although the results for user mobility and phase noise are separately evaluated, it is straightforward to envision the combining effect due to their independence.

\section{Conclusions}

This letter analyzed the impact of channel aging on the performance of zero-forcing precoding in cell-free massive MIMO, where the exchange of channel state information and precoded data via a fronthaul network raises a considerable delay. Numerical results revealed that both user mobility and phase noise affect the performance remarkably, while the aging effect due to user mobility is more severe. Consequently, the effect of channel aging should be seriously considered during the design of zero-forcing precoding  and delay-tolerant schemes such as channel prediction \cite{Ref_myOJCOMSCSIprediciton} are necessary for the practical deployment of cell-free massive MIMO.

\bibliographystyle{IEEEtran}
\bibliography{IEEEabrv,Ref_COML}

\end{document}